\documentclass[12pt]{iopart}

\usepackage{iopams}
\usepackage{amssymb}
\usepackage{bm}
\usepackage{amsfonts}
\usepackage{graphicx,epsfig}
\usepackage{graphicx,color}

\begin{document}

\title{Nonclassical correlated optical multistability at low photon level for cavity electromagnetically induced transparency}

\author{Jing Tang  \& Yuangang Deng}
\address{Guangdong Provincial Key Laboratory of Quantum Metrology and Sensing $\&$ School of Physics and Astronomy, Sun Yat-Sen University (Zhuhai Campus), Zhuhai 519082, People's Republic of China}
\ead{dengyg3@mail.sysu.edu.cn}
\vspace{10pt}


\begin{abstract}
We study the nonequilibrium dynamic behaviors in a driven-dissipative single-atom cavity electromagnetically induced transparency. The optical bistability and multistability beyond a Kerr nonlinearity are observed utilizing the optical Stark shift induced strong nonlinearity. We show that the nonequilibrium dynamical phase transition between bistability and multistability is highly tunable by the system parameters in a large parameter region. The first-order dissipative optical bistability (multistability) always corresponds to the photon-bunching quantum statistics, which indicates that the quantum fluctuations and correlations play important roles in nonequilibrium dynamics.
Interestingly, bistability and multistability with photon-bunching quantum statistics occurring at extremely low steady-state cavity photon number are observed, even under a very strong cavity driven field. Furthermore, we demonstrate that the unique cavity steady-state solution of the full quantum calculation is excellently consistent with the lowest solution based on the semiclassical mean-field approach in bistability and multistability regimes when the cavity photon number is much less than unity, albeit these nonclassical quantum states should possess strong quantum fluctuations in this parameter regime. Our results pave the way to exploring nonclassical correlated optical multistability in quantum regime, which may bring exciting opportunities for potential applications from quantum information processing to quantum metrology.
\end{abstract}

\noindent{\it Keywords\/}: Cavity-EIT, Optical bistability and multistability, Stark shift, Nonclassical correlation

\section{Introduction}
Understanding and characterizing the nonequilibrium quantum phenomena emerged by competition between an external driven field and intrinsic dissipation of the systems offers a remarkable new frontier for many-body physics~\cite{PhysRevLett.111.113901,PhysRevX.7.021020,PhysRevX.5.031028,PhysRevX.7.011012,RevModPhys.87.1379}. One fascinating paradigm is to realize optical bistability (multistability) which exhibits two (more) different steady-state solutions but with the same input parameters in driven-dissipative open quantum systems~\cite{PhysRevLett.36.1135, PhysRevA.28.2569}. Inspired by the various potential applications, e.g., all optical switches~\cite{NatureMaterials655, PhysRevLett.109.223906}, transistor~\cite{Ballarini:2013aa}, and memories~\cite{PhysRevLett.113.074301, Kuramochi:2014aa, PhysRevLett.120.225301}, optical bistability (multistability) has been extensively investigated both theoretically and experimentally in a broad range of quantum systems, ranging from atom-cavity quantum electrodynamics (QEDs)~\cite{PhysRevLett.91.143904, PhysRevA.53.1812, PhysRevLett.93.213901, PhysRevLett.121.163603,PhysRevLett.103.160403} to semiconductor microcavities~\cite{PhysRevLett.118.247402, Fink:2018aa}, surface-plasmon crystals~\cite{PhysRevLett.97.057402, PhysRevLett.99.083901, Guo:17}, cavity magnonic systems~\cite{PhysRevLett.120.057202,PhysRevApplied.12.034001,PhysRevB.103.104411,PhysRevResearch.3.023126}, and optomechanical systems~\cite{PhysRevA.83.063826, PhysRevLett.104.063601, PhysRevLett.111.043603, PhysRevLett.112.076402, PhysRevA.98.063845}. In addition, the optical bistability (multistability) is a basic and ubiquitous feature of nonlinear nonequilibrium phenomenon characterized by the first-order dissipative phase transition~\cite{PhysRevA.95.012128}. In general, the emergence of nonlinear optical phenomena relies on large average photon number and strong nonlinearity in optical resonators, i.e., Kerr nonlinearity. In particular, some pioneer explorations of optical bistability at low photon level have been experimentally studied by employing collective recoil for ultracold atomic ensembles~\cite{PhysRevLett.101.063901, PhysRevLett.99.213601}.

Up to now, the typical theoretical methods for calculating the photon emissions in cavity QEDs include semiclassical mean-field approach (MFA) with neglecting the quantum fluctuations and correlations, and the full quantum treatment by solving quantum master equation (QME) with taking into account dissipations of the system~\cite{gibbs}. Recent advances demonstrated that the optical bistability with possessing a ubiquitous feature of optical hysteresis~\cite{PhysRevLett.101.266402, doi:10.1080/09500340.2010.492919,  Paraiso2010aa,PhysRevLett.106.167002, RevModPhys.85.299} is well characterized by steady-state solution based on MFA. The experimentally measured high and low branches of optical bistability depend on the initial scanning paths for hysteresis loop. On the other hand, QME always predicts a unique quantum solution for the exact steady-state density matrix~\cite{Drummond_1980}. We remark that the unique quantum solution is independent on the initial state of atom-cavity system. These two pictures can be understood that, the unique solution of QME is the result of switching between the two branches of MFA induced by fluctuations~\cite{PhysRevA.35.1729, PhysRevA.93.033824}. However, we should note that the MFA results are usually highly consistent with the QME calculation for photon blockade in cavity QEDs at a low photon number (the steady-state photon number $n_s \ll 1$)~\cite{birnbaum2005photon,Mucke10}, albeit the quantum fluctuations, at an intuitive level, can not be neglected for this nonclassical quantum state. A readily understanding is that the nonclassical correlated single-photon states possess a strong photon-antibunching quantum statistics suppressing the quantum fluctuations of particle number for cavity photons. These results indicate that quantum correlations play an important role in nonequilibrium quantum dynamical behaviors of photon emissions. Of particular interest, exploring the quantum correlations in these nonequilibrium quantum phenomena could provide new insights and understandings in novel nonlinear phenomena in the driven-dissipative quantum systems. Remarkably, the photon correlation measurements for optical bistability have achieved tremendous advances in recent experiments for driven-dissipative Rydberg gases~\cite{PhysRevX.7.021020} and fibre cavity with Kerr-type nonlinear interactions~\cite{Fink:2018aa}. Moreover, the engineering optical bistability (multistability) at low photon level with special nonclassical quantum correlations could provide promising applications in optical communications and quantum computations~\cite{Cirac97, Duan01, Fleischhauer05}.

In this work, we present an experimental scheme to realize optical bistability and multistability in a driven-dissipative single-atom cavity  electromagnetically induced transparency (EIT) system. Utilizing the optical Stark shift enhanced nonlinearity, both optical bistability and multistability are generated with very low photon number over a wide range of system parameters. The mechanism in our proposal is obviously different from the previous studies in generating bistability with large cavity photon numbers induced by Kerr-type nonlinear interactions~\cite{PhysRevA.74.035801,PhysRevA.88.043826,PhysRevA.100.053814}. We show that the photon emissions at optical bistability and multistability regions exhibit the photon-bunching statistics (second-order correlation function $g^{(2)}(0)>1$) by solving the full QME with including the dissipations of cavity and atom fields. With an increasing optical Stark shift, the unique quantum steady-state solution of full QME is in excellent agreement with the lowest steady-state solution of MFA for bistability and multistability with $n_s < 0.01 $~\cite{birnbaum2005photon,Mucke10}. Remarkably, the first-order dissipative optical bistability (multistability) with hosting the nonclassical photon-bunching statistics is essentially different from the realization of photon blockade with strong photon-antibunching statistics in single-atom cavity-EIT~\cite{Tang2020}. The related exploration will significantly enhance our understanding on novel nonlinear quantum phenomenon in nonequilibrium dynamics emerged by the quantum fluctuations and correlations~\cite{Fink:2018aa}.

\begin{figure}[ht]
\centering
\includegraphics[width=0.65\columnwidth]{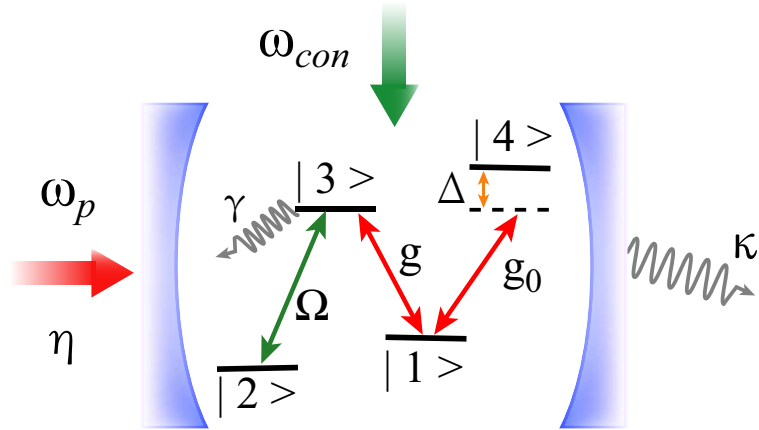}
\caption{(color online). Schematic diagram of a single $N$-type four-level atom coupled to a single-mode optical cavity. The single atom is initially prepared in populated state $|1\rangle$. The atomic transition $|1\rangle \leftrightarrow |3\rangle$ and $|1\rangle \leftrightarrow |4\rangle$ is resonantly and far-resonantly driven by the single-mode optical cavity with a single atom-cavity coupling $g$ and $g_0$, respectively. A tunable classical control field orthogonal to the cavity axis drives the atomic transition $|2\rangle \leftrightarrow |3\rangle$ resonantly  with Rabi frequency $\Omega$.} \label{scheme}
\end{figure}%

\section{Driven-Dissipative Cavity-EIT Model}

We consider a single $N$-level atom trapped inside a high-finesse cavity, as shown in Fig.~\ref{scheme}. The driven-dissipative atom-cavity system includes two hyperfine ground states $|1\rangle$ and $|2\rangle$, and two long-lived electronic orbit states $|3\rangle$ and $|4\rangle$, corresponding to the quantization axis along the cavity axis defined by a bias magnetic field ${\bf B}$. The single atom is initially prepared in populated state $|1\rangle$. As to the cavity, with bare mode frequency $\omega_c$ and cavity decay rate $\kappa$, is driven by a coherent coupling field with amplitude $\eta$ and laser frequency $\omega_p$. In our laser configuration [Fig.~\ref{scheme}], the cavity field only supports a $\sigma$-polarized photon mode orthogonal to ${\bf B}$-field originating from the vacuum-stimulated Raman transition~\cite{Tang2020}. Specifically, the cavity field is coupled to the resonant atomic transition $|1\rangle \leftrightarrow |3\rangle$ with single atom-cavity coupling strength $g$, and the far-resonant atomic transition $|1\rangle \leftrightarrow |4\rangle$ with atom-cavity coupling strength $g_0$ and a large atom-cavity detuning $\Delta$. To generate the effective Raman coupling, the atomic transition $|2\rangle \leftrightarrow |3\rangle$ is resonantly pumped by a classical control field with Rabi frequency $\Omega$.

In a large atom-cavity detuning limit, i.e., $|g_0/\Delta|\ll1$, the atomic excited state $|4\rangle$ can be adiabatically eliminated. Under the rotating wave approximation, the relevant atom-cavity Hamiltonian is given by
\begin{eqnarray}\label{Hamiltonian}
{\cal{H}}/\hbar &= \Delta_c (\hat{a}^\dag\hat{a} +\hat{\sigma}_{33}+\hat{\sigma}_{22})+ g (\hat{a}^\dag \hat{\sigma}_{13}+\hat{a} \hat{\sigma}_{31}) \nonumber\\
 & +\Omega(\hat{\sigma}_{23}+\hat{\sigma}_{32}) + U_0 \hat{a}^\dag\hat{a} \hat{\sigma}_{11} +\eta(\hat{a}^\dag  + \hat{a}),
\end{eqnarray}
where $\hat{a}^\dag$ ($\hat{a}$) denotes the creation (annihilation) operator of the cavity mode, $\hat{\sigma}_{ij}$ denotes the atomic spin projection operator with $i, j=1, 2, 3$, $\Delta_c=\omega_c-\omega_p$ is the cavity-light detuning, and $U_0=-g_0^2/\Delta$ is the tunable optical Stark shift. We should note that the large $U_0$ can be generated by elaborately selecting a long-live electronic orbital state of $|3\rangle$ and $|4\rangle$ with $g_0/g\gg 1$~\cite{PhysRevLett.117.220401,kolkowitz2017spin,bromley2018dynamics},  the corresponding matrix element of dipole-transition $|1\rangle \leftrightarrow |3\rangle$ is much smaller than $|1\rangle \leftrightarrow |4\rangle$. More details on derivation of the model Hamiltonian and experimental feasibility can be found in our previous work of Ref.~\cite{Tang2020}.

Compared with the realized strong photon blockade at single-photon resonance with the weak cavity driven strength ($\eta/\kappa\ll1$) and moderate pump field ($\Omega/g>1$) in Ref.~\cite{Tang2020}, here we focus on the nonequilibrium dynamic behaviors of bistability and multistability in the driven-dissipative single-atom cavity EIT at the very different parameter regimes. In particular, we will demonstrate that the bistability and multistability exhibit photon-bunching quantum statistics, which is essentially different from the photon blockade with hosting photon-antibunching for cavity photon emissions. Furthermore, we check that the bistability (multistability) will disappear when the cavity-light detuning $\Delta_c$ is fixed at the single-photon resonance, corresponding to the photon-antibunching quantum statistics. We should emphasize that the additional Stark shift term in Hamiltonian (\ref{Hamiltonian}) plays an essential role in realizing optical bistability (multistability) in contrast to the typical proposal in single-atom cavity-EIT~\cite{Mucke10}, as we shall see below.

\section{Equations of motion}

For the atom-cavity EIT system introduced above, the dynamical evolutions of the system determined by atom-cavity Hamiltonian $\cal{H}$ can be depicted by Heisenberg equations
\numparts
\begin{eqnarray} \label{Heisen}
i\dot{\hat{\sigma}}_{12} &= (\Delta_c-U_0\hat{a}^\dag\hat{a})\hat{\sigma}_{12}+ \Omega\hat{\sigma}_{13}-g\hat{a}\hat{\sigma}_{32},  \\
i\dot{\hat{\sigma}}_{13} &= (\Delta_c-i\gamma_{13}/2-U_0\hat{a}^\dag\hat{a})\hat{\sigma}_{13}+g\hat{a}\hat{\sigma}_{11}+\Omega\hat{\sigma}_{12}-g\hat{a}\hat{\sigma}_{33},  \\
i\dot{\hat{a}}&= (\Delta_c+U_0-i\kappa/2)\hat{a}+g\hat{\sigma}_{13}+\eta,
\end{eqnarray}
\endnumparts
where $\gamma_{13}$  is the atomic decay rate of the excited state with neglecting the damping of the ground state. Here the quantum noise of cavity and atom fields are ignored. For a weak excitation approximation, the higher-order excitations can be neglected, then Eq.(2) can be further approximated as~\cite{Mucke10, Fleischhauer05, PhysRevLett.84.5094}
\numparts
\begin{eqnarray}
i\dot{\hat{\sigma}}_{12} &= (\Delta_c-U_0\hat{a}^\dag\hat{a})\hat{\sigma}_{12}+ \Omega\hat{\sigma}_{13},  \\
i\dot{\hat{\sigma}}_{13} &= (\Delta_c-i\gamma_{13}/2-U_0\hat{a}^\dag\hat{a})\hat{\sigma}_{13}+g\hat{a}+\Omega\hat{\sigma}_{12},   \\
i\dot{\hat{a}}&= (\Delta_c+U_0-i\kappa/2)\hat{a}+g\hat{\sigma}_{13}+\eta,
\end{eqnarray}\label{Heisenberg}
\endnumparts

In steady-state regime, the time evolutions of the operators in Eq.~(\ref{Heisenberg}) are equal to zero. Then the steady-state equations can be derived as
\numparts
\begin{eqnarray}
a_s&=\frac{-\eta-g{\sigma}_{13}}{\Delta_c+U_0-i\kappa/2},   \\
{\sigma}_{12} &= \frac{-\Omega{\sigma}_{13}}{\Delta_c-U_0|a_s|^2},   \\
{\sigma}_{13}&=\frac{-g{a}_s}{\Delta_c-i\gamma_{13}/2-U_0|a_s|^2-\frac{\Omega^2}{\Delta_c-U_0|a_s|^2}}.
\end{eqnarray}
\endnumparts
here the cavity field operator $\hat{a}$ and atom field operator $\hat{\sigma}_{ij}$ in the dynamical equations of Eq.~(\ref{Heisenberg}) are replaced by $a_s=\hat{a}$ and ${\sigma}_{ij}=\langle\hat{\sigma}_{ij}\rangle$, which ignore the quantum fluctuations and correlations of the system in semiclassical mean-field treatment.

After algebraic calculations, the steady-state intracavity photon number $n_s=|a_s|^2$ obeys the following equation
\begin{eqnarray}\label{meanfield1}
C_{5}n_s^5+C_{4}n_s^4+C_{3}n_s^3+C_{2}n_s^2+C_{1}n_s+C_0=0,
\end{eqnarray}
with the corresponding binomial coefficient $C_i$ are explicitly defined as
\begin{eqnarray}\label{meanfield2}
C_5&=\alpha U_0^4,\nonumber \\
C_4&=(2\beta-4\alpha\Delta_c) U_0^3-\eta^2U_0^4,\nonumber \\
C_3&=[2\alpha(3\Delta_c^2-\Omega^2)-6 \beta \Delta_c+\chi]U_0^2+4\eta^2\Delta_cU_0^3,\nonumber \\
C_2&=U_0(\Delta_c^2-\Omega^2)(2\beta -4\alpha\Delta_c)+(4\beta \Delta_c-2\chi) U_0\Delta_c -\eta^2U_0^2[2(3\Delta_c^2-\Omega^2)\nonumber \\&+\gamma_{13}^2/4],\nonumber \\
C_1&=\alpha(\Delta_c^2-\Omega^2)^2+ (4\eta^2\Delta_c U_0-2\beta\Delta_c)(\Delta_c^2-\Omega^2)+ (\eta^2U_0\gamma_{13}^2 /2+\chi)\Delta_c,\nonumber \\
C_0&=-\eta^2[(\Delta_c^2-\Omega^2)^2+\Delta_c^2\gamma_{13}^2/4]. \nonumber
\end{eqnarray}
where $\alpha=(\Delta_c+U_0)^2+\kappa^2/4, \beta=g^2(\Delta_c+U_0)$, and $\chi=(\Delta_c+U_0)^2\gamma_{13}^2/4+(g^2+\kappa\gamma_{13}/4)^2$ are introduced for shorthand notation. By solving Eq.~(\ref{meanfield1}) analytically, the nonequilibrium
dynamic behaviors in the driven-dissipative single-atom cavity EIT can be investigated by means of the semiclassical MFA. Clearly, in the absence of optical Stark shift with $U_0$=0, the steady-state equation in Eq.~(\ref{meanfield1}) only owns one solution with the binomial coefficients $C_5=C_4=C_3=C_2=0$. However, in the presence of $U_0$, Eq.~(\ref{meanfield1}) can access up to five solutions, which indicates that $U_0$ plays an essential role in generating optical  bistability and multistability in our system.

\section{Quantum Statistics}

Now, we investigate the quantum statistical properties of photon emissions with taking into account dissipations of the cavity and atom fields by numerically solving the quantum master equation. The full driven-dissipative dynamics of the Stark shift mediated single-atom cavity EIT is described by the Lindblad master equation
\begin{equation}\label{master equation}%
{\cal{L}}\rho = -i [\hat{H}, {\rho}] + \frac{\kappa}{2} \mathcal
{\cal{D}}[\hat{a}]\rho + \frac{\gamma_{13}}{2} \mathcal
{\cal{D}}[\hat{\sigma}_{13}]\rho + \frac{\gamma_{23}}{2} \mathcal
{\cal{D}}[\hat{\sigma}_{23}]\rho ,
\end{equation}
where ${\cal{L}}$ is Liouvillian superoperator, $\rho$ is density matrix of the system, and $\mathcal {D}[\hat{o}]\rho=2\hat{o} {\rho} \hat{o}^\dag - \hat{o}^\dag \hat{o}{\rho} - {\rho} \hat{o}^\dag \hat{o}$ denotes the general Lindblad type dissipation. Compared with the analytical steady-state results predicted by MFA, we focus intentionally on quantum steady-state solution of Eq.~(\ref{master equation}), i.e., the steady-state density matrix that satisfying ${\cal L}\rho_s = 0$. As a result, we can obtain the intracavity photon number ${n}_s = {\rm Tr}(\hat{a}^\dag \hat{a}\rho_s)$ and cavity transmission $T_a= n_s/n_0$, where $n_0=(\eta/\kappa)^2$ is the bare cavity photon number without coupling to the atom field. Note that the quantum steady-state results are independent of the initial states of atomic fields.

To further characterize the statistic properties of the photon emissions, we calculate the second-order correlation function with characterizing the isolated photon correlation

\begin{equation}
g^{(2)}(\tau)=\frac{{\rm Tr}[\hat{a}^\dagger(t)\hat{a}^\dagger(t+\tau)
\hat{a}(t+\tau)\hat{a}(t)\rho_s]}{{\rm Tr}[\hat{a}^\dagger(t)
\hat{a}(t)\rho_s]^2},
\end{equation}
which directly reducing to  zero time delay second-order correlation function $g^{(2)}(0)$ with $\tau=0$. Thus the cavity photon quantum statistics show the super-Poissonian distribution when $g^{(2)}(0)>1$, corresponding to the photon-bunching with $g^{(2)}(\tau)<g^{(2)}(0)$. We have checked that the value of $g^{(2)}(0)>1$ ($<1$) can be used to well characterize the photon-bunching (-antibunching) amplitude. Here the photon-bunching (-antibunching) quantum statistics can be further confirmed by using quantum regression theorem~\cite{Tang19}.

To gain more insight into the photon quantum statistics, the second-order correlation function can be recast into
\begin{equation}
g^{(2)}(0) = 1 + \frac{\left\langle \Delta \hat{n}^2 \right\rangle}{{\rm Tr}(\hat{a}^\dagger
\hat{a}\rho_s)^2} - \frac{1}{{\rm Tr}(\hat{a}^\dagger
\hat{a}\rho_s)}= 1 + \frac{\left\langle \Delta \hat{n}^2 \right\rangle}{n_s^2} - \frac{1}{n_s}
\label{g2}
\end{equation}
with $\left\langle \Delta \hat{n}^2 \right\rangle = {\rm Tr}(\hat{a}^\dagger\hat{a}
\hat{a}^\dagger\hat{a}\rho_s) -{\rm Tr}(\hat{a}^\dagger
\hat{a}\rho_s)^2$ denotes the variance of intracavity photon number in steady state. For a fixed steady-state photon number $n_s$, it is clear that the strong photon-bunching with $g^{(2)}(0)\gg 1$ corresponds to a large variance of photon number $\left\langle \Delta \hat{n}^2 \right\rangle$. Moreover, the photon-bunching amplitude $g^{(2)}(0)$ is monotonically growing with an increasing $\left\langle \Delta \hat{n}^2\right\rangle$. As to photon blockade, the strong photon-antibunching with $g^{(2)}(0) < 1$ corresponds to a small variance of photon number $\left\langle \Delta \hat{n}^2 \right\rangle$. Remarkably, the ideal single-photon state with  $g^{(2)}(0) \sim 0$ hosts a roughly zero quantum fluctuations of particle number, where the variance of photon number $\left\langle \Delta \hat{n}^2 \right\rangle$ can be completely suppressed even at low photon level.

In our numerical simulation, we take the cavity decay rate $\kappa=2\pi\times160$ kHz~\cite{Norciae1601231} as energy unit. The single atom-cavity coupling strength is fixed at $g/\kappa=4$, which value has been demonstrated capably in current atom-cavity QED experiments~\cite{PhysRevLett.118.263601}. By employing advantages of the energy-level structure in alkaline-earth-metal atoms~\cite{PhysRevLett.117.220401,kolkowitz2017spin,bromley2018dynamics}, the atomic spontaneous decay rate is set as $\gamma/\kappa=0.047 $ by elaborately selecting the long-lived excited state of $^3P_1$ for Sr atom~\cite{Norciae1601231}. Now, the free parameters for optical Stark shift mediated single atom-cavity EIT reduce to the cavity driven strength $\eta$, cavity-light detuning $\Delta_c$, Stark shift $U_0$, and Rabi frequency $\Omega$ of the control field.

\begin{figure}[ht]
\centering
\includegraphics[width=0.95\columnwidth]{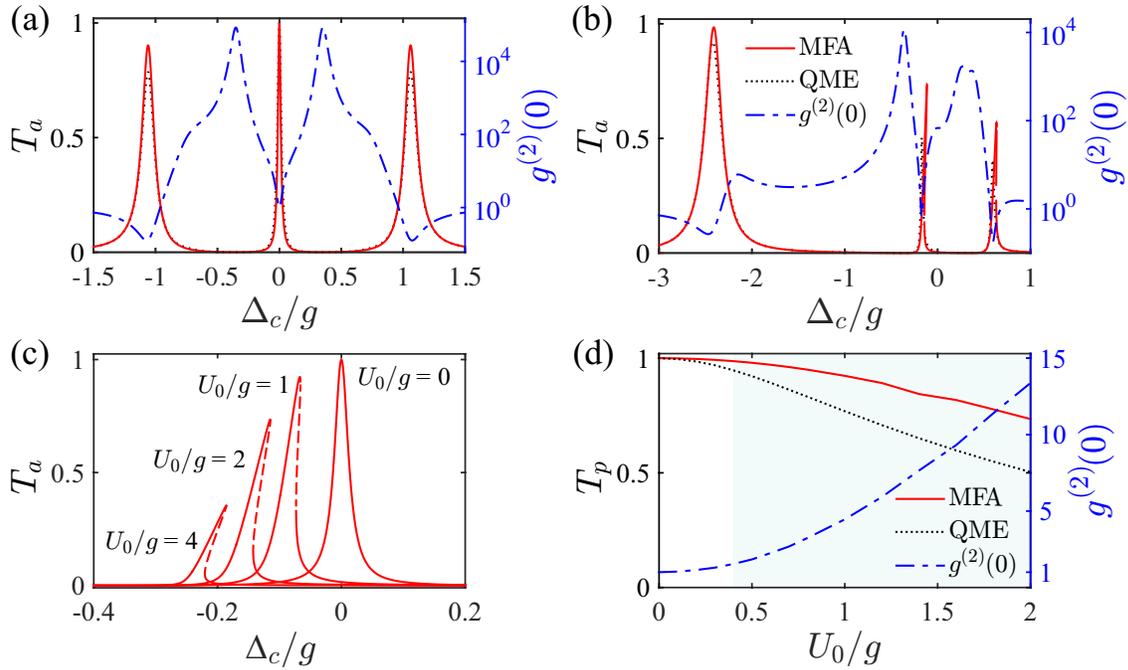}
\caption{{\color{blue}}The steady-state cavity transmission $T_a$ and second-order correlation function $g^{(2)}(0)$ (dashed-dotted line) as a function of the cavity-light detuning $\Delta_c$ for Stark shift (a) $U_0/g=0$ and (b) $U_0/g=2$  using MFA (solid line) and QME (dotted line), respectively. (c) $\Delta_c$ dependence of $T_a$ for different values of $U_0$ by using MFA, the red solid line (dashed line) represents the dynamically stable (unstable) solutions. (d) The peak vale of cavity transmission $T_p$ and second-order correlation function $g^{(2)}(0)$ (dashed-dotted line) versus $U_0$ at quasi-dark-state resonance at $\Delta_c/g\approx 0$. The light blue area denotes the bistability regime, which exhibits photon-bunching statistics with $g^{(2)}(0)>1$. In (a)-(d), the other parameters are fixed at $\eta/\kappa=0.1$ and $\Omega/g =0.35$.} \label{compare}
\end{figure}%

\section{Numerical results and discussions}

Let us first consider the weak cavity driven and classical control fields by fixing $\eta/\kappa=0.1$ and $\Omega/g=0.35$. Figures~\ref{compare}(a) and~\ref{compare}(b) display the cavity transmission $T_a$ as a function of the cavity-light detuning $\Delta_c$ based on QME (dotted line) and MFA (solid line) for Stark shift $U_0/g$=0 and $U_0/g$=2, respectively. In absence of the Stark shift with $U_0/g$=0, a narrow EIT transmission window is observed, indicating the existence of the coherent dark state at single-photon resonance of $\Delta_c=0$ with second-order correlation function $g^{(2)}(0)=1$ [dashed-dotted line in  Fig.~\ref{compare}(a)]. Besides the high EIT transmission window, the transmission spectrum also exhibits a symmetric frequency shift of the vacuum-Rabi splitting at the red and blue sidebands with $\Delta_c=\pm \sqrt{g^2+\kappa^2}$, respectively. Obviously, the transmission spectrum based on the analytical MFA  treatment is well in agreement with the steady-state result obtained by  numerically solving the master equation (\ref{master equation}).

As to nonzero $U_0$, $T_a$ exhibits asymmetry structures for both QME and MFA  results, with respect to the cavity-light detuning $\Delta_c$ [Fig.~\ref{compare}(b)]. Moreover, the conventional EIT dark-state resonance deviates slightly from the normal single-photon resonance with $\Delta_c=0$ in the presence of optical Stark shift, which is called the atomic quasi-dark-state resonance in EIT~\cite{Tang2020}. Interestingly, the photon quantum statistic exhibits photon-antibunching with $g^{(2)}(0)<1$ (dashed-dotted line) at the quasi-dark-state resonance. The mechanism for realizing photon blockade is ascribed to the optical Stark shift enhanced spectrum anharmonicity in cavity-EIT. With the increase of $U_0$, frequency shift of the quasi-dark-state resonance moves along the blue detuning of $\Delta_c$ gradually, as shown in Fig.~\ref{compare}(c). In particular, the  MFA transmission spectrum could exhibit a bistable state with an increasing $U_0$, which has a significant deviation from the full quantum treatment with a unique solution. It is clear that the hysteresis loop feature of the optical bistability  become more and more obvious with the increase of $U_0$ due to the optical Stark shift enhanced nonlinearity. The apparent  deviation between MFA and QME demonstrates that the quantum fluctuations and correlations play an important role in the bistable regime. Moreover, the peak vale of $T_p=\max[T_a(\Delta_c)]$ around the atomic quasi-dark-state resonance for MFA (solid line) is larger than the result of QME (dotted line), as well as $T_p$  gradually decreases with the increasing of $U_0$, which is due to the deviation of EIT dark-state resonance, as displayed in Fig.~\ref{compare}(d).

To address the nonclassical properties of the first-order dissipative bistable phase transition, we plot the second-order correlation function $g^{(2)}(0)$ at the quasi-dark-state resonance with respect to $U_0$ [Fig.~\ref{compare}(d)]. As can be seen, the photon emission for bistability regime hosts a photon-bunching statistic with $g^{(2)}(0)>1$, which demonstrates that the bistable state has a large quantum fluctuations than the coherent state in our driven-dissipative single-atom cavity EIT. In contrast with the photon blockade with photon-antibunching statistics, the optical bistability with featuring the photon-bunching signal for cavity emissions is clearly observed.

\begin{figure}[ht]
 \centering
\includegraphics[width=0.65\columnwidth]{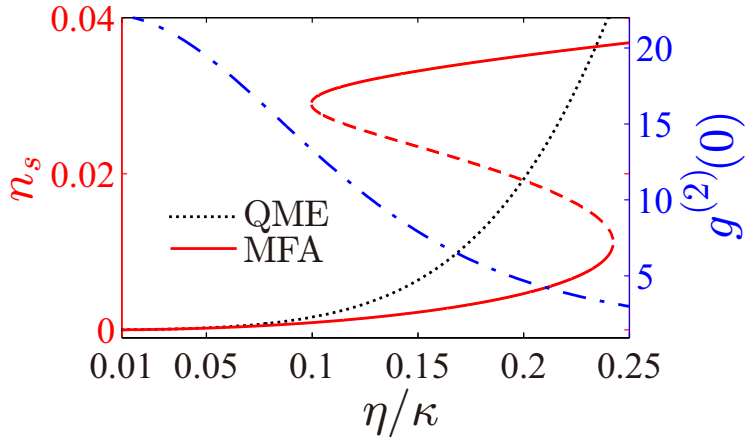}
\caption{ The intracavity photon number $n_s$ and second-order correlation function $g^{(2)}(0)$ (dashed-dotted line) versus cavity driven amplitude $\eta$ for $U_0/g=2$, $\Delta_c/g=-0.12$, and $\Omega/g =0.35$. The solid line (dashed line) corresponds to the dynamically stable (unstable) solutions using MFA  and the dotted line denotes the result of the unique solution of QME.} \label{ns-eta-Omega}
\end{figure}%

Figure~\ref{ns-eta-Omega} shows the typical $S$-shaped behavior of the optical bistability (red line) with MFA as a function of the driven amplitude $\eta$ near the quasi-dark-sate resonance with $\Delta_c/g=-0.12$ and $U_0/g=2$. We should emphasize that the light-cavity detuning is deviated from the single-photon resonance for exploring optical nonlinear phenomenon beyond the regime of photon blockade with photon-antibunching statistics. The bistability with two stable solutions and one dynamic unstable solution (dashed line) is clearly observed. In contrast to the bistability with MFA, the full quantum treatment by solving the master equation with taking into account the dissipations of the system predicts only a unique solution for QME (dotted line), without any signs of the underlying bistability. The intuitive understanding is that the unique value of QME is the weighted average of the two steady-state solutions of MFA, in which the photon emission value of QME is obviously between the two stable solutions of MFA in the bistable regime. In fact, the contradiction between QME and MFA can be reconciled by investigating the quantum nonequilibrium dynamics in the driven-dissipative quantum system as shown in Ref.~\cite{PhysRevLett.118.247402}.

In addition, the photon quantum statistics of the intracavity occupation number exhibit the photon-bunching at the bistable regime, albeit the bunching amplitude $g^{(2)}(0)$ is gradually decreasing with the increasing of $\eta$. More importantly, the steady-state intracavity photon number $n_s$ is very small in the bistable regime, i.e., $n_s\le0.04$, which indicates that the optical bistability can be generated in the quantum regime with nonclassical correlations. We should note that the photon emissions of QME are highly consistent with the MFA results even for the photon-bunching statistics with $g^{(2)}(0)> 1$, corresponding to the intracavity photon numbe $n_s\ll 1$ for $\eta/\kappa <0.1$.

\begin{figure}[ht]
 \centering
\includegraphics[width=0.75\columnwidth]{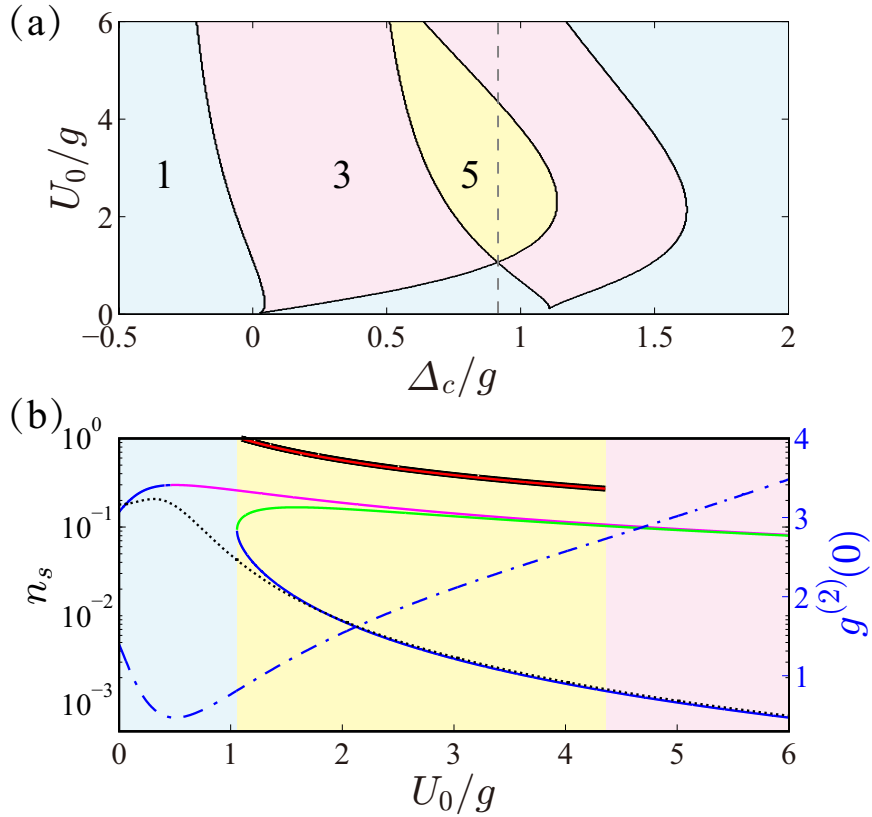}
\caption{(a) The phase diagram of the system as a function of $\Delta_c$ and $U_0$ for different numbers of steady-state solutions with $\eta/\kappa=0.6$. The differentiated light-blue, light-red, and  light-yellow regimes represent the normal steady-state (one solution), bistability (three solutions), and multistability (five solutions), respectively. The dashed line corresponding to $\Delta_c/g=0.92$ is a guide for the eyes. (b) The mean-field intracavity photon number $n_s$ and second-order correlation function $g^{(2)}(0)$ (dashed-dotted line) versus the Stark shift $U_0$ for steady-state solutions with $\Delta_c/g=0.92$. The other parameter in (a) and (b) is fixed at $\Omega/g=0.35$. 
} \label{Delta_U0}
\end{figure}%

To further investigate the nonequilibrium quantum phenomena in the system, the phase diagram for identifying different numbers of steady-state solutions is mapped in the $\Delta_c$-$U_0$ parameter plane with fixing $\eta/\kappa=0.6$, as displayed in Fig.~\ref{Delta_U0}(a). It reveals that the bistability behavior with three steady-state solutions exists in a large parameter regime (light-red regime). Besides the emergence of optical bistability, the number of steady-state solutions can be more than three, corresponding to the existence of multistability with five steady-state solutions (light-yellow regime). Remarkably, the bistability and multistability only occur at nonzero optical Stark shift ($U_0\neq0$) in the single-atom cavity-EIT system, which demonstrates that the optical Stark shift plays an important role in enhancing the optical nonlinearity~\cite{Tang2020,Tang19}. Moreover, the dynamical characteristics of bistability to multistability phase transition can be conveniently controlled by tuning the cavity-light detuning $\Delta_c$ or Stark shift $U_0$. Indeed, we have checked that the similar behaviors are observed for negative detunings when $U_0$ changes from positive to negative values. Compared with the proposed Stark shift mediated EIT, we should note that optical bistability and multistability behaviors have been studied in a $N$-type four-level atom system~\cite{PhysRevA.74.035801}, where the predicted nonlinear phenomenon can be essentially ascribed to the emerging giant Kerr nonlinearity in cavity-EIT~\cite{Imamoglu97,schmidt1996giant}.

To gain more insight into the multistability, we plot the intracavity photon number $n_s$  as a function of $U_0$ with fixing $\Delta_c/g=0.92$, as shown in Fig.~\ref{Delta_U0} (b). As we shall see, the system exhibits typical bistability and multistability behaviors with three and five steady-state solutions by varying $U_0$. Clearly,  both the optical bistability and multistability host steady-state solutions at a very low photon level ($n_s<0.01$) over a wide range of parameter $U_0$. Further insights can be obtained by calculating the corresponding second-order correlation function $g^{(2)}(0)$ ((dashed-dotted line). As can be seen, the value of $g^{(2)}(0)$ is gradually increased from photon-antibunching ($g^{(2)}(0)<1$) to photon-bunching ($g^{(2)}(0)>1$) with the increase of $U_0$. We find that the photon quantum statistics exhibit photon-antibunching with $g^{(2)}(0)<1$ for the normal steady-state phase. The photon blockade is induced by the Stark shift enhanced  energy-spectrum anharmonicity, which could facilitate the suppression of two-photon excitation~\cite{Tang2020}. There exists an optimal value of $U_0$ for generating photon blockade as demonstrated in Ref.~\cite{Tang2020}, albeit the large $U_0$ corresponds to the large nonlinearity of the system. With the further increase of $U_0$, the system enters into the optical bistability and multistability regimes corresponding to the photon-bunching statistics, which is consistent with the results displayed in Fig.~\ref{ns-eta-Omega}.

Interestingly, the unique QME solution of the full quantum calculation (dotted line) shows excellent agreement with the lowest solution of MFA in bistability and multistability regimes, as shown in Fig.~\ref{Delta_U0}(b), albeit the quantum fluctuations can not be neglected at small photon number. This result can be physically understood by means of quantum statistics. For $n_s\ll 1$ and $g^{(2)}(0)>1$ (but not very strong photon-bunching), the  variance of photon number $\left\langle \Delta \hat{n}^2 \right\rangle$ from Eq.~(\ref{g2}) is given by
\begin{equation}
\left\langle \Delta \hat{n}^2 \right\rangle = n_s^2[g^{(2)}(0) + \frac{1}{n_s} -1]\approx n_s,
\label{nflu}
\end{equation}
which means that $\left\langle \Delta \hat{n}^2 \right\rangle$ for the nonclassical correlated bistability (multistability) with photon-bunching statistics has similar properties with the  coherent state. This interesting Eq.~(\ref{nflu}) indicates that the semiclassical MFA can be successfully used to describe the bistability (multistability) at low photon level with $n_s\ll 1$. Furthermore, the photon-bunching amplitude $g^{(2)}(0)$ is increased with enhancing the nonlinearity of cavity-EIT by tuning the optical Stark shift $U_0$. As to the experimental measurement, the first-order dissipative bistability (multistability) hosts photon-bunching statistics ($g^{(2)}(0)>1$) in contrast to the photon blockade with photon antibunching ($g^{(2)}(0)<1$), which may provide a new method for diagnosing nonequilibrium dynamics by means of quantum nondemolition measurement of photon correlations~\cite{birnbaum2005photon}.

\begin{figure}[ht]
\centering
\includegraphics[width=0.75\columnwidth]{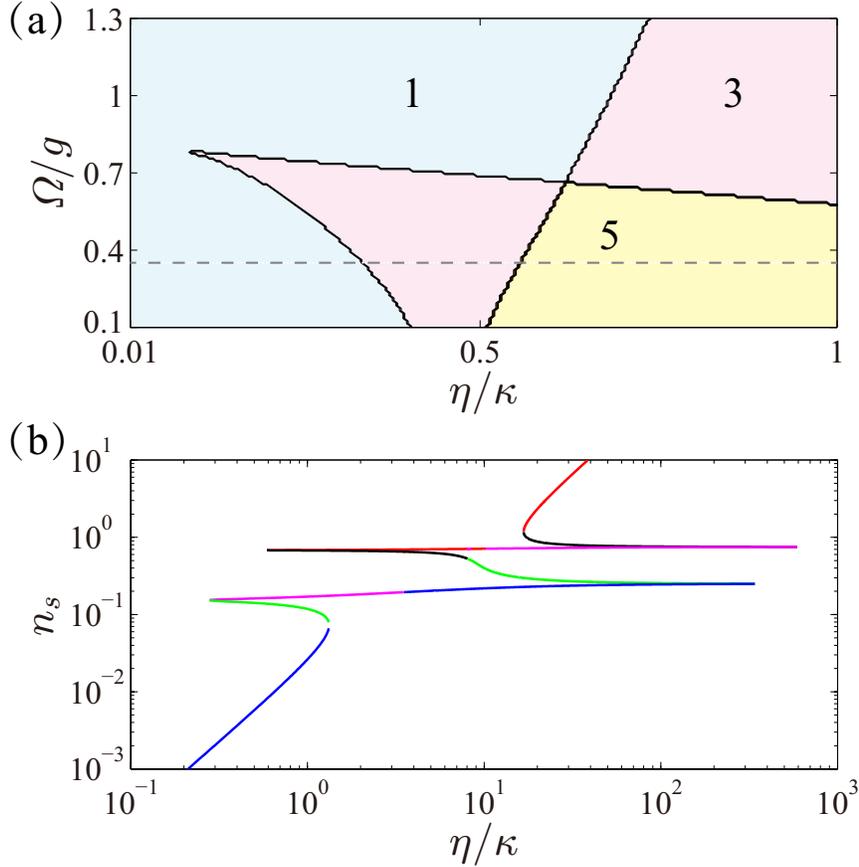}
\caption{(a) The phase diagram of the system on the $\eta$-$\Omega$ parameter plane. The differentiated regimes are labeled by numbers of the steady-state solutions. The dashed line corresponding to $\Omega/g=0.35$ is a guide for the eyes.  (b) $n_s$ versus $\eta$ for $\Omega/g=0.35$. In (a) and (b), the other parameters are $U_0/g=2$ and $\Delta_c/g=1$.} \label{eta_Omega}
\end{figure}%

Figure~\ref{eta_Omega} (a) summarizes the phase diagram of identifying different numbers of steady-state solutions on the $\eta$-$\Omega$ parameter plane with $U_0/g=2$ and $\Delta_c/g=1$. For the regime of moderate $\Omega$ and small $\eta$, the system is in the normal steady-state without emergence of bistability and multistability. In this regime, we should emphasize that the cavity emission exhibits strong photon blockade with $g^{(2)}(0)\ll 1$ as studied in our recent work ~\cite{Tang2020}. In addition, we find that the MFA result is highly consistent with the unique QME solution corresponding to the small variance of photon number $\left\langle \Delta \hat{n}^2 \right\rangle$ for strong photon-antibunching with completely suppressing the quantum fluctuations of particle number for cavity photons. In our cavity-EIT system, the multistability originates from the Stark shift enhanced nonlinearities, which occurs at a relatively small classical control field ($\Omega/g<0.7$) and moderate cavity driven field ($\eta/\kappa>0.5$). Fig.~\ref{eta_Omega} (b) shows the typical multistability behavior of $n_s$ as a function of $\eta$ with $\Omega/g=0.35$. It is clear that the system will first display bistability and then switch to multistability with the gradual increase of $\eta$. Remarkably, the predicted multistability in our model could exist over a wide range of cavity driven field $\eta$. More interestingly, for a large-scale driven amplitude, the multistability emerges with very low steady-state photon number ($n_s\sim 0.1$) even for $\eta/\kappa\gg 1$, which implies that there are exciting opportunities for applications in sensitive  all-optical switching and optical communication in quantum regime~\cite{doi:10.1063/1.4978933, PhysRevB.82.033302, Notomi:05, Amo:2010aa, PhysRevLett.108.227402, PhysRevLett.112.073901,PhysRevLett.127.133603}.

\section{Conclusions}
In conclusion, we proposed a simple experimental scheme to study the nonequilibrium dynamical behaviors in a driven-dissipative single atom placed inside an optical cavity system. It is shown that the interplay of the optical Stark shift enhanced nonlinearity and cavity-EIT gives rise to the optical bistability and multistablity. We show that the threshold and hysteresis cycles of the first-order dissipative phase transition are highly controlled by a wide range of parameters that currently available in experimental system. In particular, the observation of bistability and multistablity exhibits the nonclassical correlated photon-bunching statistics, corresponding to an extremely low steady-state photon number even for the strong cavity driven field. In addition, we find that the unique QME solution of the full quantum calculation is in excellent agreement with the lowest solution of MFA for bistability and multistability regimes, with the photon emissions exhibit nonclassical photon-bunching statistics. Remarkably, the measurement of nonclassical characteristics with photon-bunching signal for bistability and multistability could offer the opportunity to study the rich nonequilibrium dynamics through the quantum nondemolition measurement of photon quantum correlations by a Hanbury Brown and Twiss interferometer~\cite{birnbaum2005photon,PhysRevLett.118.133604}. The related exploration will further reveal new insights into the nonlinear phenomena of underlying nonequilibrium dynamics emerging by quantum fluctuations and correlations~\cite{Fink:2018aa}.

\section*{Acknowledgments}

This work was supported by the National Key R$\&$D Program of China (Grant No. 2018YFA0307500), NSFC ( Grant No. 12274473, Grant No. 11804409, Grant No. 11874433, Grant No. 12135018 ), and the Key-Area Research and Development Program of GuangDong Province under Grants No. 2019B030330001.

\section*{Data availability statement}

All data that support the findings of this study are included within the article (and any supplementary files).

\section*{References}

\end{document}